\documentclass[useAMS,usenatbib]{mn2e}

\title[Near-exponential discs]
{Near-exponential surface densities as hydrostatic, nonequilibrium profiles in galaxy discs}
\author[C. Struck and B. G. Elmegreen] 
{Curtis Struck,\thanks{E-mail: curt@iastate.edu (CS);
bge@us.watson.ibm (BGE)}$^1$
Bruce G. Elmegreen$^{2}$ \\
$^1$ Dept. of Physics and Astronomy, Iowa State Univ., Ames, IA 50011 USA\\
$^2$ IBM Research Division, T.J. Watson Research Center, 1101 Kitchawan Road, 
Yorktown Heights, NY 10598, USA}
\usepackage{amssymb}
\usepackage{amsmath}
\usepackage{graphicx}
\usepackage{multirow} 
\def\aap{{ A\&A}}

\def\aj{{AJ}}

\def\apj{{ApJ}}
\def\apjl{{ApJL}}

\def\mnras{{MNRAS}}

\def\apjs{{ApJS}}

\begin{document}
\date{\today}

\pagerange{\pageref{firstpage}--\pageref{lastpage}} \pubyear{0000}

\maketitle

\label{firstpage}
\begin{abstract}

Apparent exponential surface density profiles are nearly universal in galaxy discs across Hubble types, over a wide mass range, and a diversity of gravitational potential forms. Several processes have been found to produce exponential profiles, including the actions of bars and spirals, and clump scattering, with star scattering a common theme in these. Based on reasonable physical constraints, such as minimal entropy gradients, we propose steady state distribution functions for disc stars, applicable over a range of gravitational potentials. The resulting surface density profiles are generally a power-law term times a S\'{e}rsic-type exponential. Over a modest range of S\'{e}rsic index values, these profiles are often indistinguishable from Type I exponentials, except at the innermost radii. However, in certain parameter ranges these steady states can appear as broken, Type II or III profiles.  The corresponding velocity dispersion profiles are low order power-laws.  A chemical potential associated with scattering can help understand the effects of long range scattering. The steady profiles are found to persist through constant velocity expansions or contractions in evolving discs. The proposed distributions and profiles are simple and solve the stellar hydrodynamic equations. They may be especially relevant to thick discs, which have settled to a steady form via scattering.  
\end{abstract}

\begin{keywords}
galaxies: kinematics and dynamics---stellar dynamics
\end{keywords}

\section{Introduction}

Spiral galaxies have been known to have exponential radial profiles for a long time (\citealt{pa40}, \citealt{de59}, 
\citealt{fr70, fr07}, \citealt{va02}), with scale lengths that are independent of Hubble type for early and intermediate types \citep{de96}. The early observations did not extend to very faint surface brightnesses, nor over many scale lengths. Recent observations have gone much deeper (e.g., \citealt{bl05}  \citealt{ga09}, \citealt{er05, er08}, \citealt{po06}, \citealt{he13}, and \citealt{zh15}), showing the continuation of exponential form over about 10 scale lengths in some cases. As originally noted by \citet{fr70} the radial profiles often have a break, either turning downward in the outer parts (Type II) or upward (Type III) (also see \citealt{er05, er08}, \citealt{po06}, \citealt{he13}). However, the slope changes are often modest, and both inner and outer profiles are well fit by exponential forms. 

In addition to large disc galaxies, the surface density profiles of dwarf Irregular galaxies, which have little shear and generally no spiral waves, also follow exponential profiles out to 6 or more scale lengths (e.g., to $31\ mag\ arcsec^{-2}$ in V-band; \citealt{hu11}). The dwarfs are easily harassed by encounters with other galaxies, subject to continuing gas accretion (\citealt{va98}, \citealt{wi98}), or cycles of gas expulsion and reassertion. Thus, the stars must continuously migrate to smooth out profile disturbances, and generally do so without the aid of spirals, bars or shear. 

These observations suggest that exponentials are the generic surface density forms for the full range of two-dimensional galaxy components, and that these profiles extend over a huge range of surface brightness. They must be able to reform promptly after major disturbances, especially in dwarfs, and initially form promptly as judged by their presence in high redshift galaxies (\citet{fa12}). These more recent results greatly stress some older theories for the origin of the exponentials. 

This includes the model of \citep{me63}, based on the collapse of a uniform density, uniformly rotating sphere, with no redistribution of angular momentum. The resulting configuration has a distribution of mass as a function of angular momentum that is nearly the same as that of an exponential profile out to a radius of order 6 scale lengths. The assumptions of this simple model are questionable in light of modern disc formation models, which include processes like cold accretion from large scale structures (e.g., \citealt{ro04},  \citealt{mi12}, \citealt{br12}, \citealt{ve14}), and the exponential extent is not great enough. 

The Mestel model was updated in subsequent decades, especially to incorporate the effects of dark halos, e.g., see \citet{fa80}, \citet{mo98}, and references therein. Like the original these works consider discs that form from generally self-similar halo collapses, preserving specific angular momentum and angular momentum distributions, though some viscous redistribution was also considered. They generally assume an exponential surface density profile in the discs, and do not advance the Mestel model significantly in this regard. However, the authors do note that the model discs are close to or exceed global gravitational stability thresholds, thus providing a basic theoretical understanding for the development of massive clumps or strong waves observed in young discs (\citealt{el07}, \citealt{guo15}), and seen in more recent models (\citealt{bo07}, \citealt{ok16}). 

Another popular model suggests that viscous accretion is proportional to star formation, and then the exponential profile results from secular evolution (see \citealt{li87}, \citealt{yo89}, \citealt{fe01}, \citealt{wa09}). This type of model is discomfited by the recent observations in a couple of ways. Firstly, it requires substantial shear, which is not found in the dwarf Irregulars, and some regions of spirals. Secondly, it is not clear that disturbed exponentials can be reformed sufficiently rapidly by these processes, especially following a disturbance that does not enhance star formation.

By redistributing angular momentum and driving radial migration, bars and spiral waves can also change the surface density profile and produce exponentials. In particular, strong bars can generate double exponentials (\citealt{de06}, \citealt{fo08}). Of course, not all exponential discs have bars, including especially the dwarf Irregulars. While these several processes may play a role, they seem unable to form exponential profiles in all cases and promptly enough to account for the observations. 
 
On the other hand, simulations do show the resilience of the exponential profile in model discs. For example, the models of \citet{be15} show how steady accretion is promptly smoothed into the exponential form. \citet{el14} use analytic models to demonstrate the preservation of the exponential form when external accretion is balanced by star formation. In these models the disc can expand or shrink and the exponential scale length evolves (also see Sec. 4.1 below). 

These many observations and models urge the question of why the exponential form is ubiquitous, and promptly generated? It appears to result from a very basic and fundamental physical process. This, despite the fact that, as we will see below, it is not among the simplest possible equilibrium states. An important hint was provided by the numerical models of \citet{el13}, in which scattering clumps were introduced into simple test particle discs with an initially flat profile. In that work it was found that the scattering off the clumps generally resulted in exponential profiles. (We note that this scattering is rather different than that originating near the corotation radius of a spiral pattern discussed by \citet{se02}, \citet{ros12} and the radial evolution driven by accretion flows modeled in \citet{be15}.) This overall result did not depend on many details, such as the number or mass of the clumps, or the form of the gravitational potential, though other properties, such as the profile evolution time can depend on such parameters. 

Scattering is common to many, if not all, of the models above, especially those involving bars, waves or clumps. If scattering is indeed the underlying process that generates the exponential profiles, then the question remains of why that profile is the universal result? As does the question of the origin of the different kinds of profile (e.g., broken)? And finally, the question of how much can we learn from analytic models versus numerical simulations? Near equilibrium hydrodynamical processes can often be approximated analytically, but classical scattering problems are usually treated statistically, though with less information obtainable by analytic means. Nonetheless, biased scattering, like that proposed by \citet{el16}, can be viewed as a kind of flow to a steady state.

In the following we will demonstrate that, at least in the cases where certain, reasonable assumptions are satisfied, families of near exponentials are steady states over the range of potentials relevant to galaxy discs. Although these steady solutions are not unique, they are the simplest forms that satisfy the physical constraints. In \citet{el16} it was shown with simple scattering models (henceforth `hopping models') that the nature of disk density profiles depends on the type of bias in the scattering. This result suggests that the details of the scattering kinetics determine which of the hydrostatic solutions apply to given galaxy discs. Ultimately, the detailed study of the specific scattering processes that generate near exponential profiles in different disc evolutionary histories requires self-consistent numerical simulations.  

In the next section we describe general approximations to the hydrodynamic (Jeans) equations used to reduce these to a radial hydrostatic equation, and how scattering can be incorporated into the stellar distribution function as a Gibbs chemical potential. This formalism provides background and context, but may be skipped by readers wanting to proceed directly to discussions of solutions of the two-dimensional continuity equation \eqref{eq1} in the following sections. Section 3 discusses additional physical constraints on the solutions to the hydrostatic equation, and the nature of the constrained solutions. This discussion is expanded with a presentation of variable scalings and profile examples in Sec. 4. The hydrostatic forms derived here are the result of that evolution in mature stellar discs, which are likely to be thick discs (due to vertical scattering) in most cases. In the models of \citet{el13, el16} it was shown that it takes some time for steady, near exponential profiles to develop via scattering. We do not study that evolution in any detail here, but simple self-similar profile evolution is discussed in Sec. 5. Section 4 described some steady broken, exponential profiles, and Sec. 6 reviews the more general causes of breaks, and their evolution in simple scattering models. A summary is provided by Sec. 7. 

\section{Static profiles}
\subsection{The hydrostatic equation}

A steady state profile satisfies the time independent Jeans' equations of stellar hydrodynamics. For simplicity we will assume that the problem is essentially two-dimensional, and that the dynamics of the dimension perpendicular to the disc can be decoupled. We will further neglect azimuthal variations, and consider the cylindrically symmetric Jeans equation for radial variations. Finally, since we are looking for steady states, we assume, for the present, that not only time derivatives, but also the radial velocities are very small across the disc. In this case all terms of the mass continuity equation are negligible. We are left with the radial hydrostatic equation, 

\begin{equation}
\label{eq1}
\frac{d\Phi}{dr} - \frac{v_{\theta}^2}{r} = \frac{-1}{\Sigma} \frac{\partial}{\partial r} 
\left( \Sigma \sigma ^2 \right) = - \frac{\partial \sigma ^2}{\partial r}
- \frac{ \sigma ^2}{\Sigma} \frac{\partial \Sigma}{\partial r},
\end{equation}

\noindent where $\Phi(r)$ is the gravitational potential, $\Sigma (r)$ is the stellar surface density, $v_{\theta}$ is the local mean azimuthal velocity, and $\sigma (r)$ is the local velocity dispersion. 

In this hydrostatic equation we have used the usual pressure term for a gas. This is potentially problematic for several reasons. The first is that the gas of stars is essentially collisionless, with significant Knudsen number (i.e., significant mean free paths), whereas pressure is conventionally mediated by microscopic particle collisions. A second difficulty is that stars scattered onto eccentric orbits traverse a number of annular zones, while pressure is usually viewed as a local interaction between adjacent gas elements. This second does not seem to be a fundamental problem, e.g., eccentric orbits could be apportioned into different annuli. Moreover, the most eccentric orbits spend much time at their orbital apo-center. 

The particular environment of a clumpy disc provides a rather special solution to the first difficulty, star-surrogate scattering. That is, averaged over many scattering events, clump-star scattering can have the same effects as star-star scattering, especially in driving the system towards a steady state. For example, stars will be pushed out of over-dense regions more strongly than average regions, and the opposite in under-dense regions. This does assume that there are always clumps available to mediate pressure-like effects, an approximation that is aided by the fact that clumps can have long-range effects. Although its limitations should be kept in mind, we will adopt this approximation for the pressure.

Another simplification of equation \eqref{eq1} is the use of a single velocity dispersion in the pressure term, rather than separate radial and azimuthal dispersions. This is only valid if the dispersion is isotropic, or if the two dispersions scale by a constant anisotropy factor throughout the disc. In the clump scattering models of \citet{el13}, with the clumps placed in random locations across the disc, the velocity dispersions generated by scattering are approximately isotropic. On the contrary, there is little reason to expect scattering by asymmetric structures like bars or spirals are isotropic, except perhaps, averaged over long timescales. However, according to the recent GAIA-ESO survey results the radial and azimuthal dispersions in the solar neighborhood also do not differ greatly (\citealt{gui15}, see their Fig. 11). Of course, population kinematics can be studied in much more detail in the solar neighborhood than in external galaxies, e.g., the phenomenon of asymmetric drift (see \citealt{bt08}). Such details are not the subject of the present paper, which focusses on the general structure of steady discs, so for simplicity we will adopt the isotropy assumption throughout. The equations could be readily generalized with the inclusion of an anisotropy parameter.

If the azimuthal velocity profile, and the gravitational potential in the equation above are known, then the Jeans equation can be viewed as one equation for two unknowns, $\Sigma$ and $\sigma$. Conventionally, we might assume a polytropic equation of state, or specifically, an isothermal condition. However, the exponential profile is not a two-dimensional polytrope. Alternately, we can adopt the exponential profile, and see what velocity dispersion may support it. However, for a wide range of power-law potentials this yields an expression for the velocity dispersion squared that is an infinite power-series, which does not always converge.

\subsection{Maxwell-Boltzmann distributions}

We expect that an equilibrium profile consistent with the physical constraints will have a Maxwell-Boltzmann type distribution function. This seems reasonable since the stellar disc is unlikely to be degenerate in phase space, and we are supposing it is in a relatively relaxed state (not a \citet{ly67} violent relaxation distribution). Specifically, we assume that the (planar) distribution function is of the general form,

\begin{equation}
\label{eq2}
f = f_o e^{-\beta(\epsilon + \alpha)} ,\ 
\epsilon = \frac{1}{2} v^2 + \frac{1}{2} v_{\theta}^2 + \Phi ,\ 
\beta = \sigma^{-2},
\end{equation}

\noindent where $\epsilon$ is the specific energy, $v_{\theta}$ is the local mean azimuthal velocity, $v$ is the velocity variable in the local co-rotating frame, and $\Phi (r)$ is the gravitational potential. The square of the velocity, $v$ is the sum of the squares of the radial velocity and the difference between the azimuthal velocity and the local mean azimuthal velocity. This is analogous to the derivation of the Schwarzschild distribution function \citep[Sec. 4.4.3]{bt08}.

That derivation, however, assumes the conservation of disc angular momentum. In the present problem angular momentum is generally exchanged with the scattering centers, and the net amount in the stellar ensemble changes. This angular momentum transfer process has been well studied in the case of spiral waves, see e.g., \citet{zh98}, \citet{bt08}. In the clump scattering models of \citet{el13}, following a prompt initial drop in global angular momentum of the stars, there follows a steady, linear decrease.  Thus, we cannot simply adopt the Schwarzschild function, and instead include the $\alpha$ term in equation \eqref{eq2}.

The function $\alpha$ is a form of the Gibbs chemical potential or Fermi energy used in many areas of physics. The chemical potential is most often associated with the free energy in systems with reacting particle species. The present use is more similar to the chemical potential of free electrons in solid state systems, i.e., the energy change associated with the addition or subtraction of particles. Here the change in particle number would be due to scattering in or out of a local annulus. The free energy is important because it is the quantity available to change the state (e.g., density profile) of the system. In a disc with a large quantity of ordered motion, and little thermal energy, the free energy is large.

In the usual derivation of a distribution function like that given in equation \eqref{eq2}, via the minimization of a Boltzmann $H$ or entropy function, the velocity dispersion (or temperature) and the chemical potential are equivalent to Lagrange multipliers, and are constants (e.g., \citealt{bt08}). If we consider the disc to be a collection of radial annuli, each of small radial thickness, then each annulus can be viewed as approximately a local system, with its equilibrium distribution function described by specific values of $\alpha$ and $\beta$. 

The stellar surface density is obtained by integrating over the distribution of the two-dimensional velocity space (assuming isotropy in the local co-rotating frame),

\begin{multline}
\label{eq3}
\Sigma = 2 \int_0^{\infty} f\ 2\pi vdv \\
= 4 \pi f_o \int_0^{\infty} exp
\left( -\left[ \frac{v^2 + v_{\theta}^2 + 2\Phi- 2\alpha (r)}{2 \sigma^2} \right]  \right)
vdv.
\end{multline}

\noindent The last three terms in the exponential above depend only on radius, and so can be pulled out of the integral, and the remaining function can then be integrated,

\begin{multline}
\label{eq4}
\Sigma = 4 \pi f_o e^{-\left( \frac{v_{\theta}^2 + 2\Phi- 2\alpha (r)}{2 \sigma^2} \right)}
 \int_0^{\infty} exp \left( -\frac{v^2 }{2 \sigma^2} \right) vdv \\
= 4 \pi f_o\ \sigma^2\ 
e^{-\left(  \frac{v_{\theta}^2 + 2\Phi- 2\alpha (r)}{2 \sigma^2} \right)},
\end{multline}

\noindent which shows the dependence of $\Sigma$ on the Gibbs scattering potential. Using this last form we obtain,

\begin{equation}
\label{eq5}
\frac{1}{\Sigma} \frac{d \Sigma}{dr} = \frac{1}{\sigma^2} 
\left[ \frac{d \sigma^2}{dr} - \frac{d \Phi}{dr} + \frac{\Phi}{\sigma^2} \frac{d \sigma^2}{dr}
- \frac{1}{2} \frac{dv_{\theta}^2}{dr} + \frac{v_{\theta}^2}{2 \sigma^2} \frac{d \sigma^2}{dr}
+ \frac{d \alpha}{dr} \right] ,
\end{equation}

\noindent which can then be substituted into the hydrostatic equation \eqref{eq1}, to yield, 

\begin{equation}
\label{eq6}
0 = -2 \frac{d \sigma^2}{dr} - \frac{\Phi}{\sigma^2} \frac{d \sigma^2}{dr}
+ \frac{1}{2} \frac{dv_{\theta}^2}{dr} - \frac{v_{\theta}^2}{2 \sigma^2} \frac{d \sigma^2}{dr}
-  \frac{d \alpha}{dr} + \frac{v_{\theta}^2}{r}.
\end{equation}

\noindent This is still one equation for at least two unknowns, $\alpha$ and $\sigma$. Having rejected (globally)  isothermal and simple polytropic approximations, we turn instead to symmetry and scaling arguments, and minimization of the entropy gradient.

\section{constrained steady solutions}
\subsection{Physical Constraints}

In the most general circumstances there are five variables in the problem ($\Sigma, \sigma, \Phi, v_{\theta}$, and $\alpha$), the possible solutions are not very constrained by the radial, hydrostatic, Jeans equation alone. However, we are primarily interested in solutions generated by scattering processes. These will generally smooth inhomogeneities in phase space, so we can limit consideration to solutions that are smooth across the disc. Similarly, we do not expect the solutions to be characterized by any fixed scale lengths or wavelengths, though there will be evolving scale factors in power-law or exponential solutions. In this context, it is useful to approximate the gravitational potential as a power-law. Rotation curves and other observations suggest that over substantial ranges of radius single power-law forms are quite reasonable, and sums of power-laws can be made arbitrarily accurate. We adopt the form,

\begin{equation}
\label{eq7}
\Phi = (-1)^j  \mu r^m.
\end{equation}

\noindent  $j = 0$ or $1$, if m is positive or negative, respectively. Although this restriction on the potential will prove convenient later, equations \eqref{eq1} or \eqref{eq6} still have too many unknowns for a unique solution. The monotonic character and smoothness of this potential suggest that the forms of the other variables will be monotonic. We focus on such solutions.

In the hydrostatic equilibrium equation \eqref{eq1} gravity is balanced by both the centrifugal acceleration and the pressure gradient. We will call the pressure-balanced fraction of the gravity $\chi(r)$, so the centrifugal acceleration is, 

\begin{equation}
\label{eq8}
\frac{v_{\theta}^2}{r} = (1-\chi)  \frac{d\Phi}{dr},
\ \ \chi(r) = \chi_1 (r/a)^q, 
\end{equation}

\noindent  The second equation assumes that the function $\chi(r)$ can be approximated as a power-law, with a normalization constant $a$ and magnitude $\chi_1$. This seems a reasonable approximation, at least over limited radial ranges, especially since we are interested in smooth, large-scale structures, rather than local (e.g., wave) structures. 

There are also qualitative reasons to expect the exponent $q$ to be of small magnitude in many cases, so $\chi$ is slowly varying or nearly constant. For example, suppose that scattering centers are spread uniformly across a disc, and that their masses are distributed over a small range with a mean value such that they scatter stars that are relatively nearby, and the scattering amplitude is moderate. With these approximations we would expect the effects to be primarily local, and not extend over a large range of radius. If the scattering centers do not evolve or spiral inward too rapidly, we would expect a steady conversion of nearly circular orbital energy (centrifugal term) into the random component (pressure term). With the assumed uniform distribution of homogeneous scattering centers, a constant fraction of orbital energy should be converted into thermal energy in a given time in annuli at all radii. Thus, in this idealized case, $\chi$ would be constant across the disc, and the pressure term will scale with the centrifugal and gravitational accelerations. 

This example may be quite realistic in many cases, though numerical simulations that isolate the assumed effects are needed for confirmation. In other cases, e.g., those with a few massive scattering clumps which scatter stars through large angles, we could expect more smoothing. In such cases, the pressure term would be flatter than the gravity, and the exponent $q$ would have a positive value. If the scattering centers decreased in number or efficiency with radius, $q$ could have a negative value.

While we seek steady profile solutions, these are not true thermodynamic equilibrium states, which would be globally isothermal. Rather, we assume that fast relaxation processes have been completed, and only slow changes due to slow scattering remain. This is analogous to the case of glasses, which as a result of fast quenching do not reach their crystalline equilibrium states. Under these circumstances basic thermodynamic relations should be approximately satisfied, and gradients should be minimized by the rapid relaxation processes. For example, the fundamental relation of thermodynamics can be written in gradient form as (see e.g., Sec. 7.3 of \citealt{ha94}),

\begin{equation}
\label{eq8a}
T \frac{dS}{dr} = \frac{dE}{dr} + P\frac{d}{dr} \left( \frac{1}{\Sigma} \right)
+ \frac{d\alpha}{dr}, 
\end{equation}

\noindent where $S$ is the entropy, $E$ the internal energy, $P$ the pressure, and we have included the free energy term. We have not included a gravitational potential gradient term on the assumption that the gradients above are determined primarily by local scattering, rather than long range scattering which would sample significant changes in the fixed halo potential.

In this expression we can substitute the following, $kT = \sigma^2$, $P = \Sigma\sigma^2$, and $E = c\sigma^2$, with appropriate constant $c$. Given the assumed cold state of the disk at the onset of scattering, the entropy gradient should be small compared to the free energy gradient.  If we assume that the entropy gradient is negligible, then the previous equation reduces to,

\begin{equation}
\label{eq8b}
\frac{d\alpha}{dr} = \sigma^2 \frac{dln\Sigma}{dr} - c\frac{d\sigma^2}{dr}.
\end{equation}

\noindent This equation can be used in equation \eqref{eq5}, where the $\Sigma$ terms would cancel, yielding an equation for $\sigma$ in terms of $\Phi$ and $v_\theta$ or $\chi$. If $\chi$ is a power-law as assumed, then the general solution of equation \eqref{eq5} for $\sigma$ will also be a power-law. Equation \eqref{eq8b} shows that $\alpha (r)$ will  be a power-law as well if the surface density profile is a power-law, exponential or combination.  

\subsection{Density Profiles}

With the constraints of equations \eqref{eq7}, \eqref{eq8} and the assumed smoothness and monotonicity, we can consider several types of solution to the radial hydrostatic equation for $\Sigma$. The first type is when the velocity dispersion $\sigma^2$ scales as the gravitational potential $\Phi$. In this case, $\partial ln( \Sigma )/\partial r = -1/r$, and $\Sigma \sim 1/r$. This is not the surface density profile observed in stellar discs, and the dispersion gradient seems relatively steep compared to the observations. 

If the gravitational potential gradient minus the centrifugal term is not balanced by the dispersion gradient term (in equation \eqref{eq1}), then both must be balanced by the surface density gradient term, which leads to a couple of more types of solution. To begin, we note that if the centrifugally unbalanced gravity is zero, then, $\Sigma \sim 1/\sigma^2$. Thus, a general solution can be written $\Sigma \sim f(r)/\sigma^2$. With this form, the unbalanced gravitational term equals $-\sigma^2 dln(f)/dr$. The gravitational term and $\chi$ are both power-laws, so we might expect that $f(r)$ is as well. However, the logarithmic derivative would then yield a $1/r$ factor, reducing this to the previous case. This objection carries over to any finite series of power-law terms for $f(r)$, and many familiar transcendental functions would not satisfy the equation. 

In fact, the general solution in this case is that $f(r)$ is an exponential function. Specifically, $f(r) = \Sigma_o \sigma_o^2 exp(-h(r))$, where $\Sigma_o$ is a constant surface density. Then the pressure term equals $-\sigma^2 dh/dr$. This allows a range of power-law forms for the dispersion and the $h$ function, as long as the product scales correctly. Specifically, the following general form is consistent with a power-law gravitational potential, and the stability constraints assumed above, 

\begin{equation}
\label{eq9}
\Sigma \sim \left( \frac{1}{\sigma^2} \right) exp\left( -(r/a)^p \right),
\end{equation}

\noindent with,

\begin{equation}
\label{eq10}
\sigma^2 \left( \frac{r}{a} \right) ^{p-1} \sim \chi \frac{d \Phi}{dr} 
\sim \left( \frac{r}{a} \right)^{m+q-1}.
\end{equation}

\noindent The surface density profile in equation \eqref{eq9} is essentially a S\'{e}rsic profile with index $p$ that has a simple exponential term when $p=1$, and for example, a Gaussian term when $p = 2$. The value of $p$ is not constrained by any of the assumptions above. This key factor is evidently determined by the dynamical processes that drive discs to equilibrium states, e.g., scattering processes. An understanding of these particular processes is needed to determine $p$ (and $q$), and specifically, why a value of about 1 is realized in different types of galaxy discs. In the next two sections we will consider some special cases of the general solution above. 

To summarize the various considerations of this section, we began with five radially dependent variables ($\Sigma, \sigma, \Phi, v_{\theta}$, and $\alpha$). In the context of galaxy discs, we assumed a power-law form for the (halo) gravitational potential $\Phi$, and for $\chi$ or $v_\theta$. We used the thermodynamic relation to eliminate $\alpha$. Physical and smoothness criteria suggest that $\Sigma(r)$ is a power-law times an exponential (or S\'ersic) function. Then, the radial hydrostatic equation is solved with $\sigma$ and $\alpha$ as power-law functions, like $\Phi$ and $\chi$. These profile solutions are not unique in all cases, but appear to be the general form of monotonic solutions in a power-law potential with a slowly varying (power-law) centrifugal imbalance $\chi$. This surface density profile is clearly more general than a simple exponential; in the next section we consider when these profiles might look like observed profiles, and elaborate on the scalings of the other variables. 

\section{simple near-exponential surface densities}

In the previous section we showed that, under certain smoothing approximations, the general surface density profile in a static disc in a power-law potential is a power-law times the exponential of a power-law in radius. In this section we will examine such profiles in more detail. We will emphasize cases where the centrifugal imbalance $\chi$ is constant or nearly so. In such cases the general form for the surface density is a power-law (i.e., $\sim \sigma^{-2}$) times the simple (or nearly simple) exponential.  Observations favor such solutions, so we begin this section with some thoughts on why this might be the case. As in the discussion of $\chi(r)$ above, these considerations centre around the kinetics of scattering processes which may be responsible for the steady disc profiles. 

First consider the case of a small value of $p \ll 1$, which implies a relatively flat surface density profile. Numerical scattering models show that the latter can be achieved on long timescales after many (moderate) scatterings of a typical disc star. To achieve it on a shorter timescale requires relatively frequent long-range scattering events. Moreover, if a non-negligible surface density is to be retained despite such events, then either a barrier or a potential well sufficient to limit expansion is needed. Low values of $p$ correspond to S\'{e}rsic models of classical bulges and elliptical galaxies (like the famous $r^{1/4}$ law), rather than discs. Long-range scattering was likely associated with their formation in relatively deep halo potential wells. Strong scattering environments will also produce three-dimensional structures, rather than discs. 

Next, consider large values of $p \gg 1$, which imply rapid surface density falloffs. \citet{ko16} propose a Gaussian, $p = 2$, profile form for diffuse dwarf galaxies. These objects may be cases of weak scattering with steep initial conditions, and also a massive confining potential. Thus it appears that a reason that $p \simeq 1$ in discs is that the clumps and density waves formed within them provide the right level of moderately strong scattering to produce that form. 

Moreover, the biased hopping models of \citet{el16} produce exponential forms ($p \simeq 1$). It was suggested in that work that the necessary bias would occur naturally via clump scattering of stars formed in nearly circular orbits to more eccentric orbits. The idea is that whether scattered to elliptical orbits with larger or smaller semi-major axes there would generally be a decrease in orbital angular momentum. Then some part of the orbit would lie at radii smaller than the initial, while only a fraction of the orbits would explore larger radii, yielding the bias. Thus, we reiterate the point that kinetic scattering processes likely determine the exact form of the near-exponential density profiles, and select specific members of families of possible hydrostatic solutions. 

These arguments are qualitative, and do not constrain $p$ to be exactly unity. We will see in the following that this exponent can vary some ways from unity without destroying the exponential appearance of the profile, especially in the case of broken profiles. Indeed, we will show that such profiles can be the result of $p$ values greater or less than unity. 

\subsection{Scalings}

 In this section we will examine further the structural variables, and find some interesting scaling relations for comparison to observation. To begin, we can adopt the following specific form for the velocity dispersion described in more general terms in equation \eqref{eq10}, 

\begin{equation}
\label{eq11}
\sigma^2 = (-1)^j a \eta \chi_1 \left( \frac{r}{a} \right)^{q+1-p}  \frac{d \Phi}{dr},
\end{equation}

\noindent where $a, \eta$ are a normalization factor (as above) and a dispersion scale factor. We again assume $\chi(r) = \chi_1 (r/a)^q$, and that $j $ is defined as in equation \eqref{eq7}. By combining equations \eqref{eq7}, \eqref{eq8}, and \eqref{eq11} we can derive and tabulate a number of relations between variables, some of which will help simplify the hydrostatic equation \eqref{eq6}. To begin, we have, 

\begin{multline} 
\label{eq12}
\sigma^2 = a^m \eta \mu \chi_1 \left( \frac{r}{a} \right)^{m-1+q'}, \\
\frac{d \sigma^2}{dr} = a^{m-1} \eta \mu (m-1+q') \chi_1 \left( \frac{r}{a} \right)^{m-2+q'}\\
= (m-1+q') \frac{\sigma^2}{r}.
\end{multline}

\noindent where, $q' = q + 1 - p$. Then substitution into equation \eqref{eq8} yields,

\begin{multline}
\label{eq13}
v_{\theta}^2 = (1-\chi) \mu m r^{m}
= a^m \left[1-\chi_1 \left( \frac{r}{a} \right)^{q} \right] \mu m 
\left( \frac{r}{a} \right)^m  \\
= \frac{m}{\eta} \left( \frac{r}{a} \right) \sigma_1^2 - 
\frac{m}{\eta} \left( \frac{r}{a} \right)^p \sigma^2,
\end{multline}

\noindent where $\sigma_1^2 = a^m \mu \eta (r/a)^{m-1}$. And then, 

\begin{multline}
\label{eq14}
\frac{dv_{\theta}^2}{dr} = m^2 \mu a^{m-1} \left( \frac{r}{a}\right)^{m-1} - 
\mu m (m+q) \chi_1 a^{m-1} \left( \frac{r}{a}\right)^{m-1 + q} \\
= \frac{m^2}{a \eta} \sigma_1^2 - \frac{m(m+q)}{a \eta}
\left( \frac{r}{a}\right)^{p-1} \sigma^2.
\end{multline}

The expressions above can be used in the hydrostatic equation to get an expression for the surface density, we find, 

\begin{multline}
\label{eq17}
\frac{d ln\Sigma}{dr} = -\frac{1}{\sigma^2} \frac{d \sigma^2}{dr}
- \frac{\chi}{\sigma^2} \frac{d \Phi}{dr} \\
= - (m-1+q') \frac{1}{r} - \frac{(-1)^j m}{a \eta} 
\left( \frac{r}{a} \right) ^{p-1},
\end{multline}

\noindent which can be readily integrated,

\begin{multline}
\label{eq18}
\frac{\Sigma}{\Sigma_1} 
= \left( \frac{r}{r_1} \right) ^{-(m+q-p)} \\ 
\times exp \left[ \frac{-(-1)^j m}{p \eta} 
\left( \frac{r_1}{a} \right) ^p
\left( \left( \frac{r}{r_1} \right) ^p -1 \right) \right].
\end{multline}

This result both agrees with and gives specific form to the scaling results of equations \eqref{eq9} and \eqref{eq10}. Its derivation emphasizes that it is not an assumed form; it is derived from the assumed power-law form of the potential and the centrifugal imbalance, $\chi$. (Note that the centrifugal imbalance factor $\chi_1$ cancels out of this profile expression.) These forms are quite general and could fit a wide variety of smooth distributions over reasonable radial ranges. Next, we consider examples with various specific forms of the potential and dispersion profile, relevant to important classes of galaxy discs.

\subsection{Examples in various potentials}

 As the potential exponent $m$ ranges between values of $-1$ and $0$ the corresponding rotation curves go from Keplerian to flat, and are generally declining, so we will call them DRC cases.  The rising rotation curve cases (RRC) consist of potentials with exponents ranging from $0$ to $2$ or slightly greater. In the examples in this section we specialize (with little loss of generality) to the case where $r_1 = a$, so equation \eqref{eq18} simplifies somewhat to, 
 
\begin{equation}
\label{eq19}
\frac{\Sigma}{\Sigma_1} 
= \left( \frac{r}{a} \right) ^{-(m+q-p)} \\ 
\times exp \left[ \frac{-(-1)^j m}{p \eta} 
\left( \left( \frac{r}{a} \right) ^p -1 \right) \right].
\end{equation}

 \noindent Note also that we set $j = 0$ in the RRC case and $j = 1$ in the DRC case, as needed to obtain declining density profiles. 

\begin{figure}
\centerline{
\includegraphics[scale=0.36]{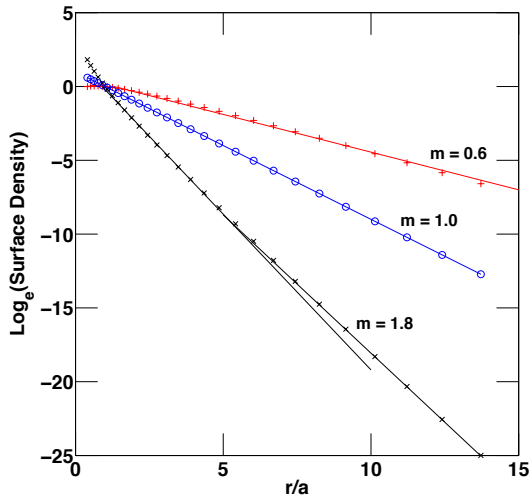}}
\caption{Dimensionless surface density profiles, $\Sigma'(r) = \Sigma(r)/\Sigma(r_1)$, in three rising rotation curve cases, as given equation \eqref{eq19} with potential exponents $m$ as labeled. Straight guide lines highlight the deviations of the profiles from pure exponential forms, and show the pure exponential form of the $m = 1.0$ case. See text for details.}
\end{figure}
 
\begin{figure}
\centerline{
\includegraphics[scale=0.36]{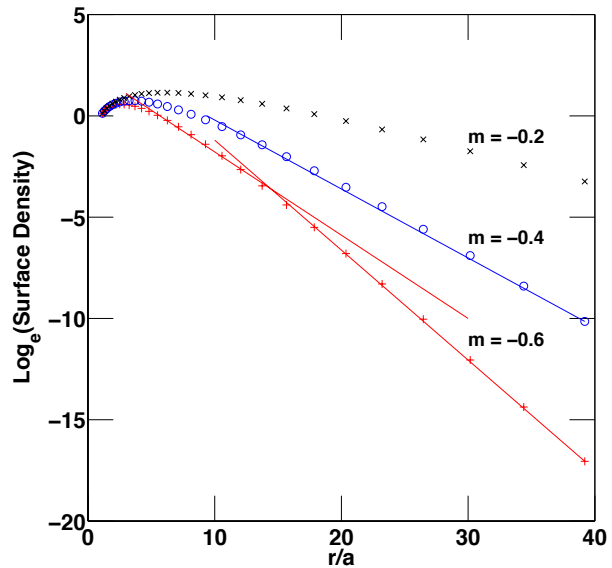}}
\caption{Surface density profiles, as in Fig. 1, but for falling rotation curve solutions of equation (19), with gravitational potential index values $m$ as indicated. The upper two curves are of nearly single exponential form into an apparent core radius, which is nearly ten times the value of $a$. The lower curve is clearly of Type II form.}
\end{figure}

\subsubsection{Examples with a simple exponential term}

 Fig. 1 shows sample RRC surface density profiles given by equation \eqref{eq19}, and Fig. 2 shows DRC cases. In these first examples we also specialize to cases with $\eta = 1, p = 1$, and $q = 0$, i.e., constant centrifugal balance. Also shown in Figs. 1 and 2 are line segments illustrating local slopes and slope changes. The first obvious feature of Figs. 1 and 2 is that despite the power-law term in equation \eqref{eq19} all of the profiles are well fit by a single or double exponential outside the core region. In particular, the RRC examples in Fig. 1 are generally very close to single exponentials, though the lowest ($m = 1.8$) case, does have a slight Type III form. In Fig. 2 the profiles could all be described as Type I or II outside the core. The guidelines on the $m = -0.6$ curve show the Type II form explicitly. Thus, regardless of the value of the potential index $m$, equation \eqref{eq19} gives exponential forms like those observed. 
 
 The steepness of the profiles outside the core is determined by the value of $m/a^p$ (see eqs. \eqref{eq17}, \eqref{eq18}). The exponent of the power-law term in equation \eqref{eq19}, $-(m+q-p)$, or in these examples $-(m-p)$ determines the profile form within the core, and to a lesser degree immediately outside the core. Observational profile decompositions have often assumed a pure exponential continuation of the disc profile into the galaxy center, and that deviations are due to other components, though more general S\'{e}rsic forms are coming into use. Various approaches can be seen in e.g., \citet{ga09}, \citet{si11}, \citet{ke12}, \citet{la12}, \citet{mu15}, \citet{sa16}. The upturns and downturns of the profile in the core generated by the power-law term suggest that using pure exponential profiles is not always correct. There may be greater or lessor disc contribution in the center depending on the gravitational potential, and the indices $p, q$, which depend on the dynamical and scattering history. The form and strength of the power-law term also affect the apparent size of the disc core. E.g., in Fig. 2 the core size appears to be about $5-10$ times the value of $a$ (or $r_1 = a$ here). These factors complicate the definition of galaxy cores, and for that matter, of bulges in late-type galaxies. 
 
 The second point, that the power-law has some effect on the profile at intermediate radii is most evident in Fig. 2, where the profile that has the clearest Type II form is the one where the exponent $-(m-p)$ is largest, so the power-law is the most nonlinear. Since $p = 1$ in these examples, any change in exponential slope is not due to the exponential term in equation \eqref{eq19}. 
 
 \subsubsection{Examples with more complex (S\'{e}rsic-type) exponential terms}
 
 Fig. 3 shows example profiles that are like those in Fig. 1 (RRC cases), but now with some different values of the exponent $p$ (but still with $\eta = 1, q = 0$). These profiles are still well fit by double exponential functions over a significant range of values of $p$, despite the nonlinearity introduced into the exponential term of equation \eqref{eq19}. However, the bottom curve in Fig. 3 pushes a bit beyond this range; the profile curvature is sufficiently strong that it takes at least three exponential segments to fit it. Note, however, that the third segment at large radii would correspond to very low brightnesses, which would be very difficult to observe.  Some recent observational studies have gone to very faint surface brightness levels, and find S\'{e}rsic-type profiles, but there is difficulty in distinguishing disc from halo stars at such low levels (e.g., \citealt{co13}, \citealt{ds14}).

\begin{figure}
\centerline{
\includegraphics[scale=0.35]{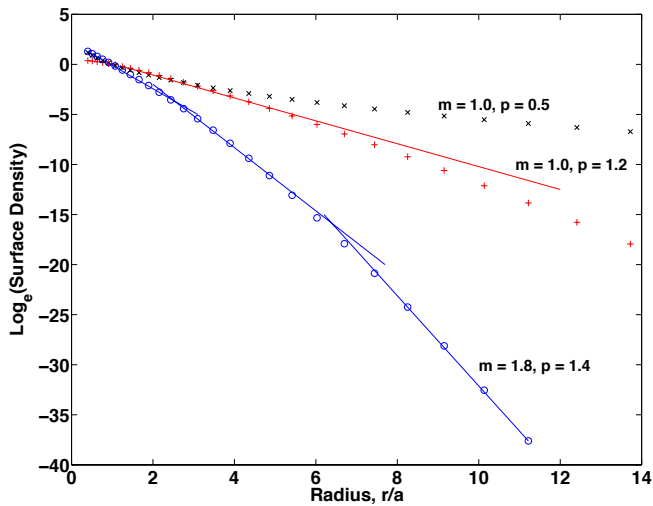}}
\caption{Surface density profiles, as in Fig. 1, but for rising rotation curve solutions of equation (19), with gravitational potential index values $m$ as indicated and various values of the exponent $p$. The upper two curves are of nearly single exponential form into an apparent core radius, which is nearly ten times the value of $a$. The lower curve is clearly of Type II. See text for details.}
\end{figure}
 
 It is clear from the lower two curves that values of $p > 1$ generate stronger downward curving profiles, including Type II profiles if the curvature is not too great. The top curve, on the other hand, shows that Type III profiles can be generated from small values of $p$. It would be harder to generate Type III profiles in FRC (flat rotation curves) cases because the more nonlinear power-law term would resist the upward turn of the profile. This is in accord with the observation that Type II profiles are more common in late-type discs. The index $p$ is the inverse of the usual S\'{e}rsic index, and the implication of Fig. 3 is that rather being pure exponentials, discs could have S\'{e}rsic-like profiles with S\'{e}rsic indices ranging from about $2/3$ up to $2$ or more. This is very much in accord with the observation that secular or pseudo bulges, believed be formed from disc instabilities, have S\'{e}rsic indices near or slightly less than $2$. Evidently, it would take only some vertical scattering to convert the low $p$ profiles considered here into secular bulges. {\it We might further conjecture that the sequence of S\'{e}rsic profiles from late-type discs through bulges to de Vaucouleurs-like forms in ellipticals, is primarily a sequence in mean scattering length and magnitude integrated over the life of the galaxy.}
 
 Velocity dispersion profiles are very modest power-laws in these models, i.e., $\sigma \sim r^{0.5(m+q-p)}$. In the cases of the nearly simple exponential profiles in the upper two curves of Fig. 1, this exponent equals 
 $-0.2$ and $0.0$. In cases like those shown in Fig. 2 it is between $-1.0$ and $-0.5$. 
 
 Of course, with the power-law term equation \eqref{eq19} is not exactly a S\'{e}rsic form. It is more like the `core S\'{e}rsic' profiles considered recently for some elliptical and bulge profiles (\citealt{gr03}, \citealt{tr04}, \citealt{sa16}), although the core function adopted in those works is not quite the same as the power-law in equation \eqref{eq19}. Nonetheless, we again see the unity of form across various galaxy components. 
 
Although Figs. 1-3 show a variety of forms, an even greater variety can be obtained by allowing variations in the parameters $q, \eta$ or $r_1$ in equation \eqref{eq18}. The sizes and shapes of cores can be changed with $q$ and $r_1$, as can the intermediate profile, and the slope of the outer exponential changed with $\eta$. Some of these adjustments overlap those that can be made with $m$ and $p$. 

Conversely, constraining all these parameters with limited observation sets will be nearly impossible. Constraining the gravitational potential and $m$ independently of stellar disc kinematics should be possible in some cases. Surface photometry to very faint levels might constrain $a$ and $p$ (via profile curvature) to some degree. However, comparisons to observation are complicated by the fact that observed profiles can be different in different wavebands, e.g., due to stellar population gradients (e.g., \citealt{ba08}, \citealt{he13}, \citealt{zh15}). Also there may be radial variations in mass-to-light ratios, so mass and light profiles may differ \citep{zh15}. Azimuthal velocity and velocity dispersion profiles will help constrain combinations of the exponents and the factor $\chi_1$. Quantities related to scattering, including $\eta, \chi_1, q$, and $p$, may be functionally related, at least for specific scattering processes. If so, numerical simulations might be used to discover these relations. Additionally, self-gravitating discs may be further constrained, i.e., via the Poisson equation. Thus, the results above provide more of a framework, rather than a fully predictive theory, of the steady outcomes of the many dynamical processes in galaxy discs.

\section{Evolving profiles and bends}
\subsection{Expanding or contracting exponentials}

In the previous section we examined the hydrostatic force balance of stellar discs consisting of annuli in local equilibria, and with zero mean radial velocities. Models, like those of \citet{el13} and \citet{bo07}, show that the evolution driven by massive clumps in the disc works to form an exponential profile regardless of the initial density distribution.  In discs with declining rotation curves stars are scattered to large radii relative to the initial disc size, and the exponential profile is eventually extended as well. This expansion is characterized by a linear mean radial velocity profile, which quickly develops and then relaxes with a gradually decreasing slope. In rising rotation curve discs there is a slow inward diffusion of stars, though the mean radial velocity is generally negligible at all radii. However, \citet{el14} found that in accreting discs with rising rotation curves there is a steady state inflow solution to the hydrodynamic equations with a linear radial velocity profile. In these cases and the outflow cases, the flow generally preserves the exponential form (but not the scale length) of the surface density profile.

This fact suggests that a disc may evolve through a series of local, near equilibrium, hydrostatic states like those described above while experiencing slow expansions or contractions. Specifically, equation \eqref{eq1} may be approximately satisfied throughout this process, with the time-dependent part of the velocity equation satisfied separately. That is, the time evolution of the velocity profile and the advection term approximately cancel each other, such that,

\begin{equation}
\label{eq20}
\frac{\partial v_r}{\partial t} + \frac{v_r}{r}  \frac{\partial }{\partial r} \left( rv_r \right) = 0.
\end{equation}

\noindent We consider a separable solution of the form,

\begin{equation}
\label{eq21}
v_r = b(t)r,
\end{equation}

\noindent where $b(t)$ is the time-dependent slope of the velocity profile. This equation can be integrated to obtain the (Lagrangian) motion of a star moving with the mean flow,

\begin{equation}
\label{eq22}
\frac{r(t)}{r(t_o)} = exp 
\left( \int_{t_o}^t b dt \right),
\end{equation}

\noindent where $t_o$ is an arbitrary initial time. With this expansion law there will be no crossing of annuli within the disc and the continuity equation will be satisfied, though annular widths will increase with the expansion. The mean mass within each disc annulus will be conserved, so the product $\Sigma r (\Delta r)$ equals a constant, and, 

\begin{equation}
\label{eq23}
\Delta r(t) = r_O - r_I = \Delta r(t_o) e^{\int bdt}
= r \left( \frac{\Delta r}{r} \right)_{t_o} \sim r,
\end{equation}

\noindent where $r_I$ and $r_O$ are the inner and outer radius of an annulus. This implies that the quantity $\Sigma r^2$ is constant in time within the expanding annulus, a result that is unique to the linear expansion law (equation \eqref{eq21}). Clearly, for this result to be consistent with maintaining an exponential surface density profile, the exponential scale length must evolve simultaneously. Conversely, for a slow expansion to maintain the near equilibrium profiles, it must also maintain the linear expansion velocity profile. 

To complete this expansion solution we substitute equation \eqref{eq21} into equation \eqref{eq20} and solve for the function $b(t)$. The result is, 

\begin{equation}
\label{eq24}
\frac{b(t)}{b_o} = \frac{1}{1 + 2 b_o t_o \left( \frac{t}{t_o} - 1 \right)}.
\end{equation}

This shows the steady decline of the slope of the velocity profile, expected from the scattering simulations. Substituting this into equation \eqref{eq22} gives the radius-time relation for an expanding annulus,

\begin{multline}
\label{eq25}
\frac{r}{r(t_o)} = exp \left[ ln \left(1 + 2 b_o t_o \left( \frac{t}{t_o} - 1 \right) \right) ^{\frac{1}{2 b_o t_o}} \right]\\
= \left( 1 + 2 b_o t_o \left( \frac{t}{t_o} - 1 \right) \right) ^{\frac{1}{2 b_o t_o}}.
\end{multline}

As long as $b_ot_o > 1/2$ this yields a moderate annulus expansion rate, and differentiation shows that this rate decreases with time. Note that the scales $a, r_1$ must also evolve in accordance with equation \eqref{eq25} to preserve the steady profiles of equations \eqref{eq18} and  \eqref{eq19}. Simple scattering models (\citealt{el13}, \citealt{el16} and additional unpublished models) show that flat or declining rotation curve discs tend to scatter stars outwards and expand, while solid body models do the opposite. Thus, the profile expands and flattens in the former case, and vice versa in the solid-body case.

\subsection{Angular Momentum Evolution}

We have noted above how the stellar ensemble generally loses angular momentum as orbits become more elliptical. However, there are other processes that change angular momentum. Scattering centers such as clumps or bars will experience dynamical friction, so their orbits will change, and angular momentum may be transferred to the ensemble of stars or the halo (see review of \citealt{se14}). Accreted material will also evolve as it settles into the disk, and exchange angular momentum with the stars. Thus, even if the stellar ensemble settles to a near steady state, it is unlikely to remain in exactly the same state in a changing disk environment. 

There are a couple of circumstances where the steady profiles described above could coexist with the processes of continuing evolution. The first is when this evolution is slow or secular, i.e., characterized by a timescale that is longer than the profile adjustment timescale of a few scattering times. 

The second circumstance is when the angular momentum exchanges drive a radial flow where the radial velocity scales with radius $r$. This is the case considered in the previous subsection, and also in \citet{el14}, and which preserves the exponential profile form. The case considered in \citet{el14} with accretion balanced by star formation is a nice example of where the near exponential profiles may exist as a quasi-equilibrium state that slowly changes because of external forces. 

\section{Breaks and bends}

If some classes of expansions, contractions and regulated accretion preserve exponential surface density profiles, other less symmetric ones might generate breaks. In fact, a number of processes have been proposed to generate profile breaks. First of all, the results of the Sec. 5.1 showed that these forms are the equilibrium profiles in certain parameter ranges. Secondly,  in many cases, Type II profiles may be the result of a gravitational potential whose form varies with radius, e.g., due to the presence of a bar in the inner disc. \citet{la14} find a large fraction of Type II profile breaks associated with rings in barred galaxies. Even in unbarred galaxies, they could be the result of a connection between the disc and the specific angular momentum of the halo according to \citet{he15a}. Mergers \citep{pe06} or stellar population gradients may also be responsible for some bent profiles, including Type IIIs \citep{yo07}. Type III profiles might also be generated by scattering off bars \citep{he15b}.

Scattering models (\citealt{el13}) suggest additional possible causes relevant to unbarred galaxies. For example, FRC discs beginning with arbitrary (e.g., flat) surface density profiles often evolve through an intermediate stage characterized by a double exponential; Fig. 4 shows an example. The outer, steeper part of the profile extends beyond the initial disc and consists of scattered particles. Later the profile form usually settles to a single exponential, which evolves to ever flatter slopes in flat rotation curve cases. In strongly scattering models this stage is brief, enough stars are soon scattered outward to fill in a single exponential profile. In moderately scattering cases it can be persistent. In the strongly scattering cases the scattering centres are likely to disappear quite rapidly, via frictional infall \citep{bo07} or dissolution. Then the intermediate, double exponential phase may get `frozen in' following their disappearance. 

The outer exponential in a Type II profile can be interpreted with equations \eqref{eq11} and \eqref{eq18}. Most of the scattered particles in that region will be on quite eccentric orbits. Their velocity dispersion relative to the local circular velocity will be larger than in the inner exponential. Note that these circular velocities cannot be equated to the local mean azimuthal velocities; they must be measured independently or determined by extrapolating the inner rotation curve. We would predict that when the outer disc has been scattered, and the orbits have been elongated, then the stellar rotation speed will be less than the gas speed. This is subject to the caveat that secondary scattering may put stars back into more circular orbits with rotation speeds comparable to the gas. The phase space volume for circular orbits is relatively small, however, so complete re-circularization is unlikely to equalize the azimuthal velocity of stars and gas in the outer disc.  

\begin{figure}
\centerline{
\includegraphics[scale=0.35]{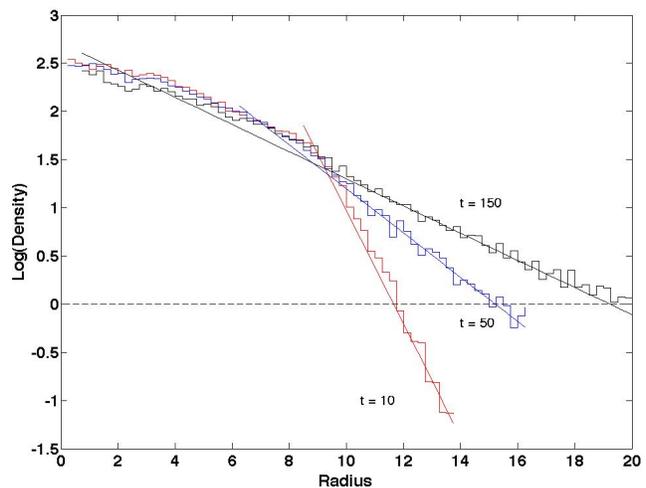}}
\caption{Surface density profiles at three times showing the evolution in simple scattering models like those described in \citet{el13}. The dimensionless units are also the same as those in that reference, e.g., the orbital period of a particle on a circular orbit of radius 1.0 is $2\pi$. The gravitational potential is that of a flat rotation curve. The line segments are least squares fits to the scattering over the range of their extension. They show that at the earliest times the profiles are approximately broken, Type II form. At the last time the profile approaches Type I form, thereafter grows flatter as it continues to expand.}
\end{figure}

There is an analogous effect producing Type III profiles in solid-body (e.g., dwarf) discs with scatterers. Profiles in these discs tend to evolve inward \citep{el13}. If the initial profile was not exponential, then the exponential tends to develop first in the inner regions, and to steepen there with time. This naturally gives something like a Type III profile, though it may take a long time to develop, unless scattering is strong. This is different than the Type III scenario proposed by \citet{mi12}. They suggested that the outer part of Type III profiles was produced by accretion, and an accompanying increase in velocity dispersion. 

In \citet{el16}, we described numerical scattering models with two stellar generations each initialized with flat surface density profiles. Profile evolution slows with time, but continues nonetheless, so the second generation was not able to catch up and match the profile of the first, though both had evolved to exponential forms. The second generation was assumed to form from a larger gas disc, and dominated at large radii. This also generated a broken surface profile, and is another natural way to produce broken profiles.

Finally, we note that the calculation of the previous section applies to bent profiles as well as pure exponentials; they too could be stretched or compressed in a self-similar manner. The evolution of bends and breaks in surface density profiles can be quite dynamic, while still tending to settle to equilibrium exponential forms. 

\section{Summary and Conclusions}

The goal of this paper has been to better understand the exponential surface brightness or surface density profiles in galaxy discs, whose phenomenology was briefly reviewed in the Introduction. Part of the mystery of exponential discs is that while they appear to be equilibrium states, they are not fit by simple, polytropic solutions of the stellar hydrodynamics equations. This paradox has been sharpened by recent observations showing that the exponential profiles can extend over many scale lengths, as shown by previous studies referenced in Sec. 1. Although the simplest solutions do not suffice, the stellar hydrodynamics or Jeans equations still provide strong constraints on the surface density and velocity dispersion profiles. In Sec. 3 we used physical constraints to narrow the range of possible solutions, and proposed specific forms. These resulted in an exponential or S\'{e}rsic-type radial dependence in the surface densities, but with an additional power-law term. This form also describes the disc hopping model of \citet{el16}. This power-law term has several effects. The first is slightly changing the exponential slope in some cases. These power-law modified profiles can be well fit by two distinct exponential segments in some parameter ranges. These fits usually resemble observed Type II and III disc profiles.  The second effect is an upturn, or downturn, of the profile at the smallest radii, which is also seen in observed profiles, but may be difficult to separate in the observations from a bulge contribution. Additionally, in-plane velocity dispersions are predicted to follow moderate power-law functions with radius. 

The adoption of equation \eqref{eq8} (with power-law $\chi$) reduces the range of steady solutions. Consider the consequences of relaxing it. With these assumptions the effective pressure, $\Sigma \sigma^2$, is an exponential in all cases, since power-law terms cancel. The pressure gradient term in equation \eqref{eq1} (pressure gradient divided by $\Sigma$), is a pure power-law in radius with the same power as the other terms in that equation. However, if the pressure was not purely exponential, then generally there would be two or more pressure terms with different radial scalings. The scaling of the centrifugal term must be altered to balance them. Observations, e.g., comparisons between gas and stellar kinematics, where the former are assumed to represent near circular orbits in centrifugal balance, tend to suggest that the gravitational and centrifugal terms do scale similarly, so such extra terms are usually small. Measurements of the in-plane velocity dispersion scalings in discs are difficult, but would be very helpful for further testing the scalings predicted above. 

In sum, while stellar discs in galaxies are not simple, cylindrically symmetric, isothermal, exponential (or polytropic) atmospheres, they come rather close. Firstly the observations showing velocity dispersions do not vary by large factors across discs. Secondly, the distribution functions and surface density profiles of Sec. 2 - 4 above are locally, but not globally, isothermal. This despite the fact that galaxy discs are very cold, and nearly in centrifugal balance, so we might not expect even local thermal relaxation. Another difference is that, because stars can be scattered over large distances, and not confined to a local annulus, the distribution functions contain a chemical potential term.  The surface density solutions generally contain a power-law dependence on radius as well as the exponential, in each case appropriate to the specific gravitational potential. 

This cored S\'{e}rsic-type profile, extending over a range of S\'{e}rsic index values from about $3/4$ to $2$ provides a unification with the equilibrium structure of bulges and ellipticals. Pseudo-bulges, in particular, are believed to have indices a bit lower than 2. The overlap with some disc profiles makes sense if the they are indeed secularly formed from discs. 

Another perspective is obtained by eliminating $r$ between equations \eqref{eq12} and \eqref{eq18}. This yields a density-pressure ($\Sigma - \sigma^2$) relation that differs from a polytropic one in several ways. Firstly, it is more complicated. Secondly, it contains the centrifugal imbalance term $\chi$. It appears that the solutions here are generalizations of the polytropes to cases with the additional effects of biased scattering over a broad range.  

Hopping models \citep{el16} and numerical scattering models \citep{el13, st16} also show that the stellar disc structure is not described by a constant entropy (polytropic) equilibrium state. A true equilibrium would be a globally isothermal structure, which could only be achieved on a long, two-body relaxation timescale. Even early-type galaxy discs retain a large kinetic energy in near circular rotation, which can be viewed as free energy that will ultimately be converted into thermal energy. This exponential structure is a flow, driven by biased scattering, with an entropy gradient. It is a slow and slowing flow, with a nearly hydrostatic structure, in which scattering minimizes the entropy gradient. Such nonequilibrium, hydrostatic structures may be useful models in a variety of other applications where scattering is important.

\section*{Acknowledgments}

We are grateful to an anonymous referee for very helpful suggestions. We acknowledge use of NASA's Astrophysics Data System, and the NASA Extragalactic Data System.

\bibliographystyle{mn2e}

\bsp
\label{lastpage}
\end{document}